# Defending Against Adversarial Attacks by Energy Storage Facility

Jiawei Li[1], Jianxiao Wang[2], Lin Chen[1*], Yang Yu[1,3]

1.Institute for Interdisciplinary Information Sciences, Tsinghua University, Beijing, China, 100084;
2.State Key Laboratory of Alternate Electrical Power System with Renewable Energy Sources, North China Electric Power University, Beijing, China, 102206;
3.Shanghai Qi Zhi Institute, Shanghai, China, 200003.

***Abstract*—Adversarial attacks on data-driven algorithms applied in the power system will be a new type of threat to grid security. Literature has demonstrated that the adversarial attack on the deep-neural network can significantly mislead the load forecast of a power system. However, it is unclear how the new type of attack impacts the operation of the grid system. In this research, we manifest that the adversarial algorithm attack induces a significant cost-increase risk which will be exacerbated by the growing penetration of intermittent renewable energy. In Texas, a 5% adversarial attack can increase the total generation cost by 17% in a quarter, which accounts for around $\$2\times10^7$. When wind-energy penetration increases to over 40%, the 5% adversarial attack will inflate the generation cost by 23%. Our research discovers a novel approach to defending against the adversarial attack: investing in the energy-storage system. All current literature focuses on developing algorithms to defend against adversarial attacks. We are the first research revealing the capability of using the facility in a physical system to defend against the adversarial algorithm attack in a system of the Internet of Things, such as a smart grid system.**

## I. Introduction

### A. Background and Motivation

Adversarial attacks on the algorithms in cyberspace threaten the operation of the physical side of the power system. The current literature in the field of deep learning manifests that the adversarial attack can fully fail a learning model by designing a negligible noise in the input data. While the smart grid is a cyber-physical system (CPS) that relies on data and learning mode, adversarial attack brings a new type of threat: attacking a physical power grid by designing noise to mislead data-driven models. For instance, attacking the load-forecast model can cause more power outages by injecting an undetectable designed noise into input data. Since a load forecast bias is very common, adversarial attacks can be disguised as normal forecast bias. This unique feature differentiates the adversarial attack from the conventional physical attack. While the adversarial attack is low-cost and hard to be detected, the attack can frequently occur and lead to a large but undetectable loss.

To resist the ever-growing threat, researchers have made significant efforts. The research into this data attack problem has generally focused on state estimation and they are closely related to the Supervisory Control and Data Acquisition (SCADA) system for improving the state detection of this system. However, the power demand side, which is not monitored by SCADA, is overlooked. Meanwhile, cyber-attacks on load demand only require knowledge of machine learning-based load forecasting methods, which can mislead load forecasts by generating adversarial samples. Actually, grid operators rely on the results of load forecasts to schedule the start-up and shutdown of generating units and the amount of electricity generated. A forecast bias in demand caused by an adversarial attack on electricity users could mislead the power dispatch and incur an economic burden. Consequently, the accuracy of the forecast load directly affects the power system's safe and stable operation and economy.

Most of the current research tries to develop algorithms to defend against adversarial attacks, such as the generative adversarial networks (GAN) model. However, few of the existing studies notice the possibility of utilizing the physical facilities to protect a system of internet-of thing (IoT) from algorithmic attacks. While the algorithmic attack causes cost by disturbing the physical system's operation, we argue that physical facilities can play a protective role. In this research, we analyze the progress of how an algorithmic attack on the cyber side leads to an economic burden on the physical side and discuss the possibility of using physical facilities in the grid, such as the energy storages, to defend against the cyber attack.

### B. Related Works

Adversarial attacks have been researched in many practical applications[1], such as computer vision[2] and natural language processing[3]. Qiu et al.[4] proposed the definition of adversarial samples. If the activation path of a sample is not included in the training activated path, then the sample is classified as an adversarial sample. There are some systematic ways to find adversarial examples or attack neural networks. Moosavi-Dezfooli et al.[5] presents the DeepFool algorithm, which is a simple but very useful algorithm to find adversarial examples or attack the neural network. Later, Madry et al.[6] uses the Projected Gradient Descent (PGD) algorithm to find adversarial examples. The algorithm runs the PGD algorithm to maximize the loss, and project back to the $\varepsilon$-ball if needed. However, the above methods are all aimed at specific scenarios and do not involve an adversarial cyber attack on the power system. Very recent literature has noticed the possible risk of adversarial attack bringing to the power grid. Chen et al.[7] found that a white-box attack can bring significant errors in the prediction of the load forecast model.

As a combination of a cyber-physical system, the security of the modern grid system is threatened by both algorithmic and physical attacks. Harvey malware [8] and load redistribution attacks [9] have been studied to force the grid operator into incorrect actions. Several methods to defend against adversarial attacks have been proposed. Mohsenian-Rad and Leon-Garcia [10] proposed a cost-efficient load protection strategy. However, it may be not efficient for many

(Corresponding author: chen_180707@163.com)

distributed IoT devices. Amini et al.[11] explained the fundamental characteristics of dynamic load-altering attacks and designed a protection system by formulating and solving a non-convex pole-placement optimization problem based on feedback from the power system frequency. Li et al.[12] evaluated the vulnerability of deep neural networks (DNN) through adversarial attacks and designed an adversarial-resilient DNN detection framework to defend against such attacks. Overall, most of the previous works are focusing on proposing the strategy and the algorithm to protect the power grid security. However, most current research just focuses on the consequence of attacking in the cyber part. It is unclear whether and how the cyber error causes realistic economic loss and physical damage. Little literature focuses on physical facilities to directly defend the cyber attacks.

### C. *Contrbution*

i) To our best, we are pioneering research using the facility of the power grid in the physical world to resist algorithmic adversarial attacks in cyberspace.

ii) We found the significant and complex impacts of the adversarial attack on the power grid and found the existence of extremely venerable hours. We applied a white-box attack to simulate an adversarial attacker to attack the load forecasting algorithm by injecting poison data.

iii) We highlight that the penetration of intermittent renewable energy will aggravate the impact of the adversarial attacks and exacerbate the vulnerability of the power grid system.

## II. Problem Formulation

In the actual dispatch operation of the power system, operators always make decisions for minimizing the overall cost, which will inevitably harm the interests of some users. Therefore, the user has the motivation to tamper with the data to gain more benefits. In this way, users are strategic and may mislead the model by uploading false data to obtain greater benefits. At the same time, energy storage (ES) can be controlled to improve the robustness of the power system, as shown in Fig. 1. Thus, how adversarial users attack existing models is a problem we need to discuss below.

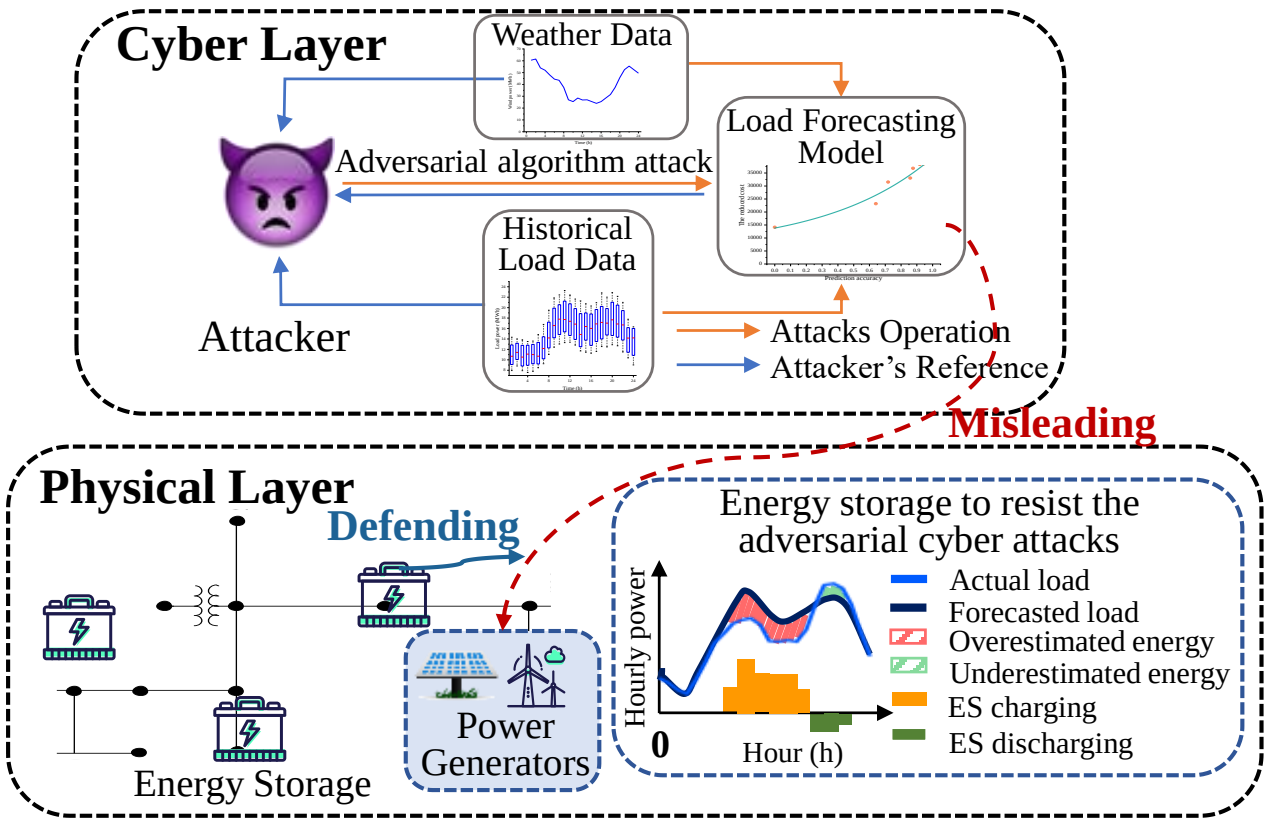

**Fig. 1. The schematic of adversarial attacks on load forecasting algorithms resulting in the threats of power system operations.**

In this section, we introduce the forecasting and attacking models according to Chen et al.[7]. We assume there exists an adversarial attacker trying to attack the load forecasting algorithm by injecting poison data, which is simulated by a white-box attack.

### A. *Load Forecasting Formulation*

Load forecasting is fundamental to the operation of power systems. Historical load data and weather data are trained by load forecasting algorithms to label electricity load and weather data at different moments. Once the weather data is known, it is possible to predict the load data for future moments. To improve the accuracy of the predictions, other data such as latitude and longitude are also fed into the training data. Therefore, the training dataset is :

$$\sigma_{t+1} := \{X_i, T_i, E_i\}_{i=t-H}^{t} , \tag{1}$$

where $H$ represents the period scale for intercepting historical data; $X_i$ is the load demand at the time $i$; $T_i$ is the temperature at time $i$; $E_i$ is the other information at the time $i$.

$f_\theta$ is defined as a demand forecasting model, and $\sigma_{t+1}$ is defined as the input data. Estimation of $\theta$ is given by minimizing the difference between model predictions and real values:

$$\min \sum_{t=H+1}^{T} L_1(f_\theta(\sigma_t), X_t), \tag{2}$$

where, $L_1$ is the loss function to measure the performance of the forecasting algorithm, which is defined by mean squared error (MSE). Here, the recurrent neural network (RNN) is used to characterize load data and then predict them.

### B. *White-Box Attack*

The purpose of adversarial attacks by strategic users is to mislead the predicted value, thereby increasing the deviation. the predicted value from the actual value. They can control the demand $X_i$, but this kind of control is limited. In addition, we consider that strategic users can also control the temperature information $T_i$ in a small range. The attacked $\vec{X}_t'=[X_{t-H}',\ldots,X_t']$, $\vec{T}_t'=[T_{t-H}',\ldots,T_t']$ and other information $\vec{E}_t=[E_{t-H},\ldots,E_t]$ together constitute the input data $\sigma_{t+1}'$. We consider the simplest goal, which is to maximize the difference between the predicted value and the true value:

$$\begin{aligned} &\max L_2(f_\theta(\sigma_{t+1}'), X_{t+1}) \\ &\text{s.t. } \|X_i' - X_i\| \le \varepsilon_1 \cdot \|X_i\| \\ &\qquad \|T_i' - T_i\| \le \varepsilon_2 \cdot \|T_i\| \\ &\qquad i = t-H, t-H+1, \ldots, t \end{aligned} \tag{3}$$

where, $\varepsilon_1$ and $\varepsilon_2$ ensure small-scale control. To avoid detection, the range of perturbation caused by the poison data is limited. $L_2$ is the loss function, defined by Mean Absolute Percentage Error (MAPE).

In this paper, we consider the scenario when the attacker knows the detail of the load forecasting algorithm and figure out the attacking method by a white-box attack approach. The attack strategy can be obtained by solving (3). Here, the PGD algorithm proposed by Madry et al. [6] is used for white box attacks to obtain adversarial examples for load prediction after an attack.

$$x_{k+1} = \prod_{B_\varepsilon(x)} [x_k + \alpha \cdot \text{sign}(\nabla_x L(\theta, x, y))], \tag{4}$$

The core of the attack method PGD is the formula (4), where $x_k$ is the historical data at $k$-th iteration and obviously

$x_0$ is the actual data; $y$ is the value to be estimated; L calculates the prediction loss based on the parameter $\theta$; $\alpha$ is the step size to update adversarial examples; $\prod B_\varepsilon(x)$ means projection to $x$'s $\varepsilon$-neighborhood. The adversarial examples are constructed by an ascent in the sign direction of the gradient several times.

In this scenario, the attacker will design the attacking strategy by minimizing the size of the perturbation with two restrictions. One is the limit of the perturbation range, the other is that the forecast is generated by the given load forecasting model. The attacker uses an adversarial neural network to figure out the optimal perturbation and inject the perturbation into the input data of the load forecasting model. Then the load forecasting model is misleading and will project a wrong prediction of the demand.

## III. System Model

In this section, we consider a simplified dispatching process, without considering transmission line constraints. In the dispatching process, there are two supply sources, i.e., renewable energy sources and traditional thermal plants to meet the requirements of load demand every time slot. Besides, there is ES to control the intermittency and uncertainty of green energy.

**Renewable Energy.** For the zero-fuel cost, renewable energy should be utilized as much as possible. There could exist time slots during which renewable energy is more than demand. We set the ratio of the remaining renewable energy to the renewable energy gap as the indicator $\rho$, which is an effective measure to assess the potential to install energy storage for the time-shifting of renewable energy.

$$\rho = \sum_{t=1}^{T} \left[ w_t - D_t \right]^+ \Big/ \sum_{t=1}^{T} \left[ D_t - w_t \right]^+ , \tag{5}$$

where, $[\cdot]^+$ is the positivity operator (max{ · , 0}); $w_t$ is the output of green energy at time $t$; $D_t$ is the demand at time $t$.

**Thermal Plant.** The unit fuel cost is assumed to be fixed. When the renewable energy output is insufficient, thermal power plants produce electricity. Since the electricity power company is a profit seeker, so cheaper electricity is generated first.

**Load Demand.** It should be forecasted in advance. The power dispatcher makes decisions on the amount of power to be generated by each generator to meet the load demand based on the load forecast. However, the load forecast will be biased after adversarial cyber attacks. When the load demand is overestimated at each time step, the operator tends to charge the excess power into the ES. If there is no surplus in ES, unfortunately, the excess power can only be wasted, which will result in excess costs. When load demand is underestimated, the ES discharges to the system and the more expensive thermal plants may output power. Therefore, reasonable control of ES can help withstand such adversarial data attacks to avoid additional costs.

The renewable energy prediction and load prediction profiles are inputted into the model. However, considering the user's adversarial attack behavior on load demand, we assume that renewable energy is predicted precisely in the existing forecasting model, and the load demand is biased. Deviations in load forecasting will cause curtailment costs and additional energy costs.

The system operator may seek to minimize the operating cost. The system model is formulated as:

$$\min C_{total} = \sum_{t=1}^{T} \begin{bmatrix} \sum_{i=1}^{N} c_t P_{i,t} + c_t \left( FD_t - D_t - g_t \right)^+ \\ + c_t^{'} \left( D_t - FD_t - b_t \right)^+ \end{bmatrix}, \tag{6}$$

where, $T$ is the total number of time slots; $N$ is the number of thermal plants; $c_t$ is the marginal cost at time $t$; $c_t^{'}$ is the highest marginal cost among thermal plants used to satisfy the demand at time $t$; $P_{i,t}$ is the actual power produced by thermal plant $i$ at time $t$; $FD_t$ is the forecasted demand at time $t$.

subject to

$$\sum_{i}^{N} P_{i,t} + w_t = D_t + g_t - b_t , \tag{7}$$

$$B_t = B_{t-1} + g_t - b_t , \tag{8}$$

$$0 \le B_t \le \overline{B}, \tag{9}$$

$$B_0 = 0, \tag{10}$$

$$0 \le P_{i,t} \le P_{i,\max} , \tag{11}$$

where, $g_t$ and $b_t$ are the charging and discharging power of the ES at time $t$, respectively; $B_t$ is the stored energy of the ES at time $t$; $B_0$ and $B_T$ are the initial and ultimate energy levels, respectively; $\overline{B}$ is the capacity limitations of the ES; $P_{i,\max}$ is the capacity of thermal plant $i$. Since ES is expensive, the capacity $\overline{B}$ is not large in fact. Therefore, $\overline{B}$ is much smaller than $D_t$.

Constraint (7) is the power balance. The energy balance equation and the capacity limitation of the ES are given by constraints (8) and (9), respectively. Constraint (10) represents that the energy storage is charged and discharged from none electricity. Constraint (11) shows the capacity limit for the thermal plant.

**Energy Storage.** In this paper, we consider an easy control strategy of ES. When the load demand is overestimated, the ES should be controlled to store the excess power (renewable power or excess power generated for over-forecasted demand) until the ES is fully charged. In the following time, the stored electricity in ES can be discharged to make up for the lack of electricity.

---

**Algorithm:** Energy Storage Control

---

**Input:** Demand $D_t$, forecasted demand $FD_t$, renewable energy $w_t$, and ES status last time slot $B_{t-1}$;
**Output:** Cost $C_{\text{total}}$ and ES status $B_t$.
**If** $w_t<FD_t$ **then**
  Discharge ES first, and then generate power by thermal plants to meet $FD_t$-$w_t$ demand;
**Else**
  Charge ES as much as possible until the ES cannot be charged or the ES is charged by $w_t$-$FD_t$;
**EndIf**
**If** $FD_t>D_t$ **then**
  Charge ES as much as possible and the rest power is wasted;
**Else**
  Utilize the rest renewable energy first, and then discharge ES as much as possible. If the demand is still not satisfied, use thermal power plants to meet it.
**EndIf**
Calculate $C_{\text{total}}$ and $B_t$ from the decision;
**return** $C_{\text{total}}$ and $B_t$

---

## IV. Case Studies

To evaluate the impact of adversarial attacks on the power system, we obtain the demand, renewable and thermal plant data from the electric reliability council of Texas (ERCOT)[13]. To maintain the same order of magnitude, we multiply the original solar data by a coefficient of 16000. And then we multiply wind and solar energy by coefficients 3, 4, 5, and 6.5 to reflect different proportions of net renewable energy, i.e., 0.25%, 3.2%, 12%, and 43%. Weather data[14] includes hourly temperature, precipitation, air density, and cloud cover in Houston in 2012. The maximum demand exceeds 80000 MW per hour. The battery capacity is 16000 MW.

### A. *Adversarial attacks' performance*

We compare the operation cost with and without adversarial attacks on the electricity pool model. We found that the adversarial attack can largely inflate the generation cost. For instance, a 5% attack on the load forecast model can increase the quarter generation cost by 200 million dollars, which is roughly 17% of the total generation cost.

To analyze the cost impact of the adversarial attacks in each hour. We define the concept of the cost-loss ratio in each hour. The cost loss ratio caused by an attack indicates the ratio of the difference between the cost of an attack ($\varepsilon$ =3%) and the cost of no attack to the average cost per hour. The attacks will bring cost losses in general, but some hours will also bring cost gains, as shown in Fig. 2. The cost loss ratio of most hours is near 0, while the cost loss ratio of some hours is greater than 0 or even less than 0. We notice that the distribution of adversarial attack's hourly impact is bimodal. Therefore, the power-system cost is extremely venerable in some hours shown as in Fig. 2. Thus, the hours are divided into three types: benefit hours (the cost-loss ratio is lower than zero), loss hours (the cost-loss ratio is higher than zero) and extremely vulnerable hours.

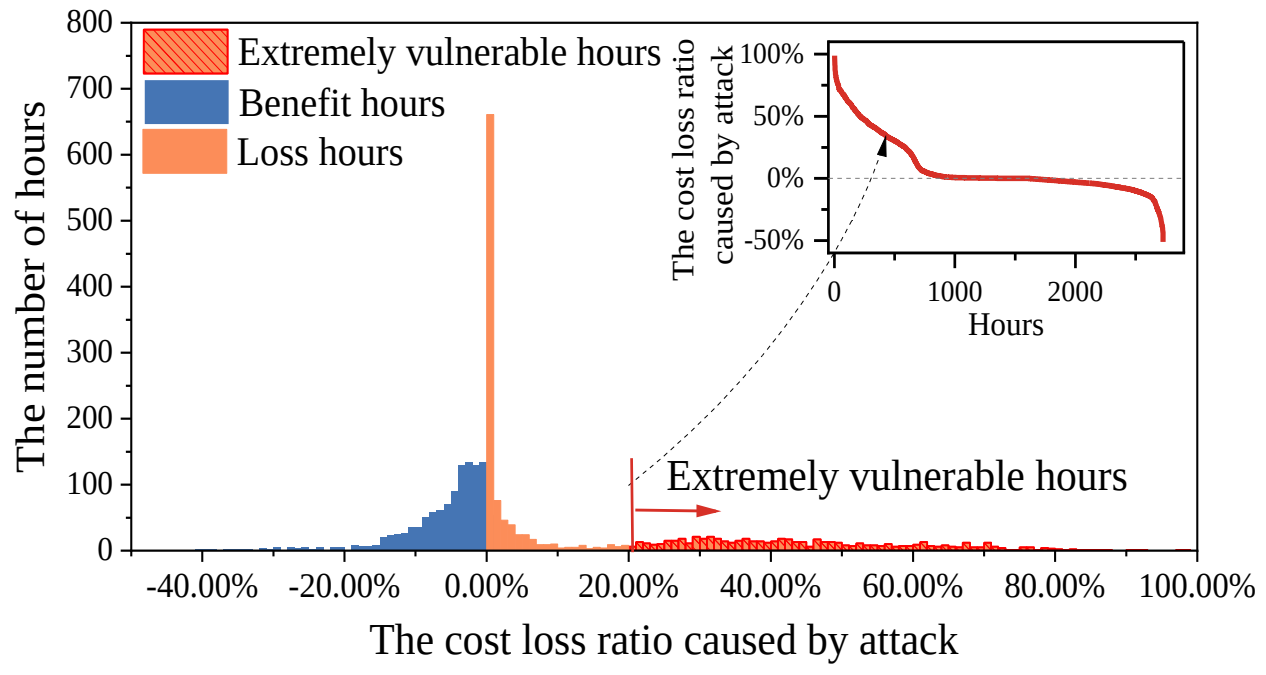

**Fig. 2. The cost loss ratio caused by attack under 3.2% proportion of renewable energy.**

We analyze the distribution characteristics of the three types of hours and summarize them in Fig. 3(a). We found that the benefit hours are mainly distributed in 7:00-8:00 from September to December, and the loss hours are distributed in 9:00-22:00 from September to October. Among them, the extremely vulnerable hours are concentrated in the daily peak of electricity consumption (11:00 to 19:00). This is because electricity consumption in peak hours is large and the electricity price is higher, which is more vulnerable to user attacks. While in November and December, the loss hours appear more disorderly, and even on a certain day in December, the whole day is extremely vulnerable hours, which relates to the high electricity consumption on that day. Therefore, the system operator of ERCOT has to particularly be careful in those extremely venerable hours.

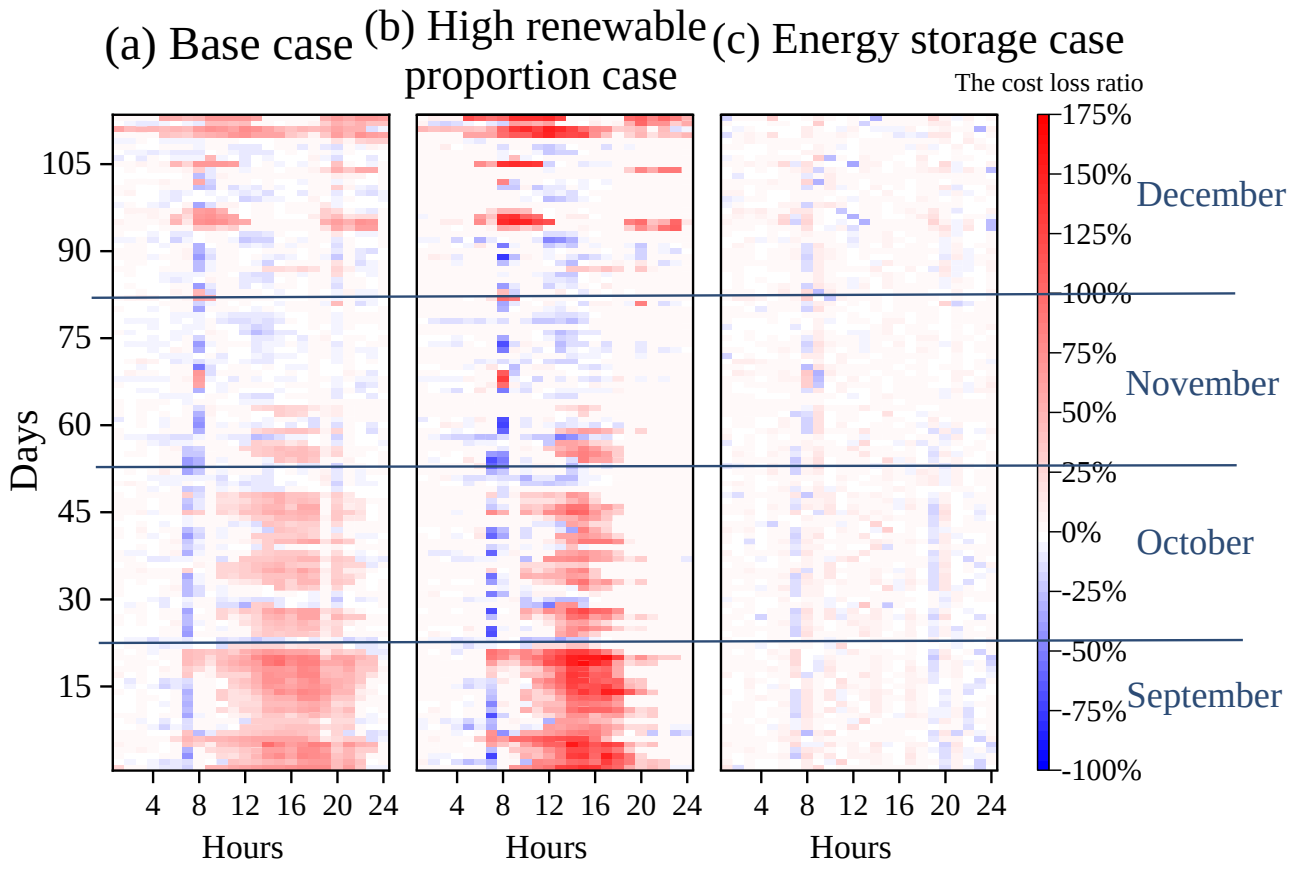

**Fig. 3. A heat map of the cost loss ratio over 24 hours from September to December in 2012. (a) The base case under an attack ($\varepsilon$=3%) with 3.2% renewable proportion. (b) High renewable proportion case under an attack ($\varepsilon$=3%) with 43% renewable proportion. (c) Energy storage case to control battery to resist attacks ($\varepsilon$=3%) with 3.2% renewable proportion.**

### B. *The impact of renewable energy*

By comparing Fig. 3(a) and Fig. 3(b), it can be seen that increasing the proportion of renewable energy will not affect the distribution of the three types of hours, but will increase the cost-benefit of the benefit hours and the cost-loss of the loss hour, respectively. The increase in the proportion of renewable energy makes the system more vulnerable to adversarial attacks, but the impact varies from hour to hour. To further explore the impact of the proportion of renewable energy on the cost loss ratio, we performed scatter plots of the cost loss ratio under different renewable energy proportions in the same hour.

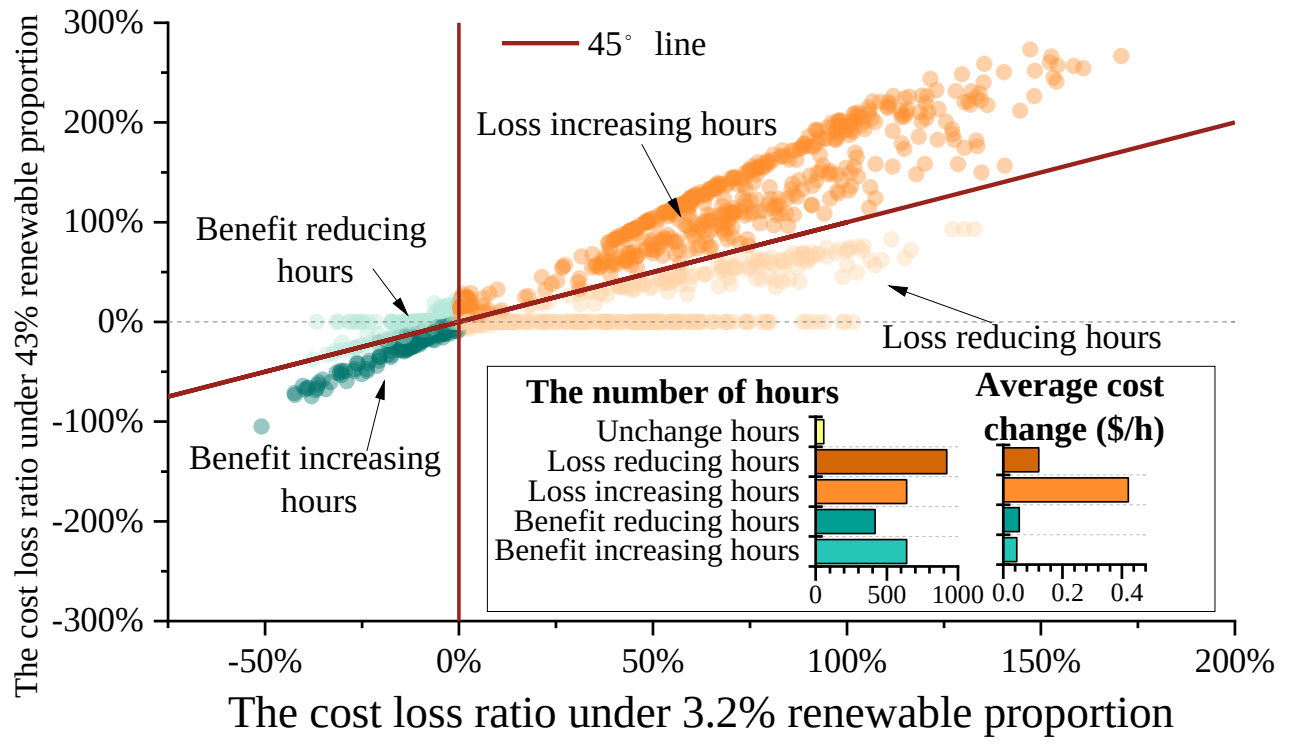

**Fig. 4. For the same hours, the cost loss ratio (between no attack and $\varepsilon$ =3% attack) under 3.2% renewable proportion is related to that under 43% renewable proportion. The equation next to the line represents the fitting function. In the lower right corner, 43% renewable proportion compared to 3.2%, '*unchange*' means that the cost loss ratio remains unchanged; '*increasing*' means the hours with increasing cost loss ratio; '*reducing*' indicates the hours with decreasing cost loss ratio.**

As shown in Fig. 4, with the increase in the proportion of renewable energy, the change costs in the benefit hours are small, while the change costs in the loss hours are large. For the benefit hours, their reducing and increasing change not much. While for the loss hours, their cost loss ratios increase the most, and the average change cost reached 0.422$/h,

which indicates the proportion of renewable energy has a greater impact on loss hours.

### *C. Energy storage to resist adversarial attacks*

To evaluate the ES's ability to resist attacks, we selected the two periods (1-500 and 1500-2000 hours) to demonstrate the charging and discharging conditions of the ES in the face of the attack, as shown in Fig. 5 (a) and (b). In 1-500 hours, the prediction error caused by the attack is alternately positive and negative, while most of the prediction errors in 1500-2000 hours are negative. In the case of alternating positive and negative errors, ES can be utilized to the maximum and the value of ES can be revealed to the greatest extent, as shown in Fig. 5 (c).

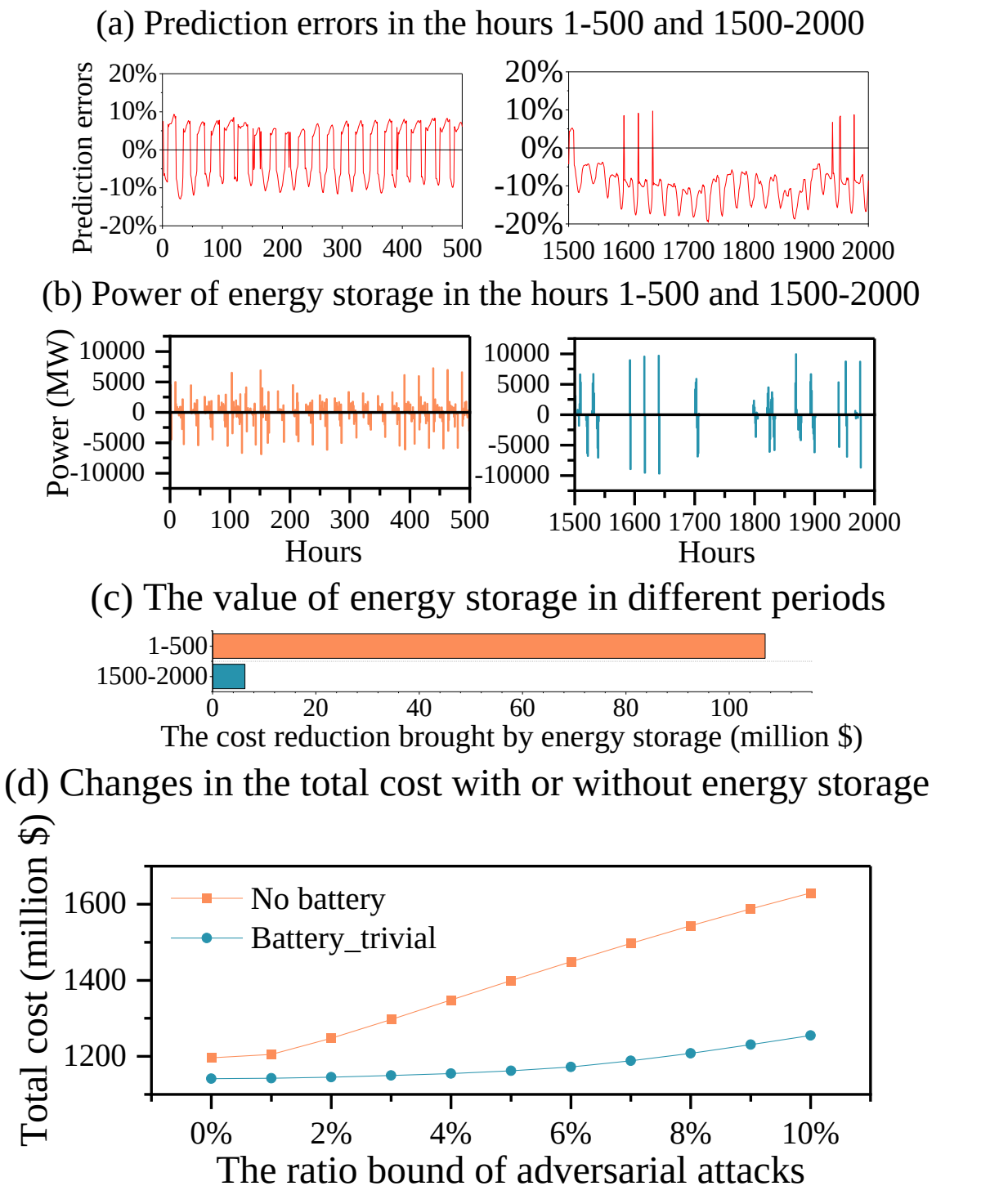

**Fig. 5. Energy storage output and total cost after the adversarial attack. (a) Actual load and forecasted load after the attack ($\varepsilon$ =3%) in the hours 1-500 and 1500-2000. (b) Charging and discharging of energy storage in the corresponding hours. (c) The cost reduction brought by energy storage in different periods. (d) Changes in the total cost with or without energy storage under varying degrees of attack.**

The changes in the total cost with or without ES under varying degrees of attack are compared in Fig. 5 (d). The total cost has risen sharply with the increase in the degree of attack without the ES, while the total cost has only slightly increased with ES. Moreover, the higher the attack degree, the greater the cost reduction brought by ES. This is because ES can effectively enhance the ability of loss hours to resist attacks. Compared to Fig. 3 (c) with Fig. 3 (a), the red part almost disappears after ES participates and all hours become unchanged hours. Therefore, ES effectively enhances the system's ability to resist attacks every hour, and ultimately brings cost reductions.

## V. Conclusion

In this research, we reveal that the adversarial attack on the load-forecast DNN model will bring significant cost inflation in the ERCOT market. However, the adversarial attack's impact varies over hours. The attack increases cost in some hours but decrease cost in others. We argue that the ERCOT has to care about venerable hours. We also found that the penetration of intermittent renewable energy will exacerbate the capability of an algorithmic adversarial attack on inflating generation cost. Thus, when the power grid is decarbonized by deepening the penetration of wind and solar power, the grid becomes more venerable to adversarial attacks. We also demonstrate the capability of using physical facilities to fight against the algorithmic adversarial attack. Thus, our research proposed a novel perspective for adversarial research in the field of power grid study and IoT research.

In our future work, two issues deserve an in-depth study: 1) The application of physical facilities to defend against cyber attacks will be extended to more practical scenarios. 2) It is worth estimating the carbon footprint brought by such adversarial attacks on the power system.

## References

[1] LeCun, Yann, Yoshua Bengio, and Geoffrey Hinton. "Deep learning." nature 521(7553: 436-444, 2015.

[2] Athanasios Voulodimos, Nikolaos Doulamis, Anastasios Doulamis, and Eftychios Protopapadakis. Deep learning for computer vision: A brief review. Computational intelligence and neuroscience, 2018, 2018.

[3] Voulodimos, Athanasios, et al. "Deep learning for computer vision: A brief review." Computational intelligence and neuroscience 2018, 2018.

[4] Qiu Yuxian, et al. "Adversarial defense through network profiling based path extraction." Proceedings of the IEEE/CVF Conference on Computer Vision and Pattern Recognition, 2019.

[5] Moosavi-Dezfooli, Seyed-Mohsen, Alhussein Fawzi, and Pascal Frossard. "Deepfool: a simple and accurate method to fool deep neural networks." Proceedings of the IEEE conference on computer vision and pattern recognition, 2016.

[6] Madry, Aleksander, et al. "Towards deep learning models resistant to adversarial attacks." arXiv preprint arXiv:1706.06083, 2017.

[7] Chen Yize, Yushi Tan, and Baosen Zhang. "Exploiting vulnerabilities of load forecasting through adversarial attacks." Proceedings of the Tenth ACM International Conference on Future Energy Systems, pages 1–11, 2019.

[8] L. Garcia, F. Brasser, M. H. Cintuglu, A.-R. Sadeghi, O. Mohammed and S. A. Zonouz, "Hey my malware knows physics! attacking PLCs with physical model aware rootkit", Proc. Netw. Distrib. Syst. Secur. Symp., 2017.

[9] Y. Yuan, Z. Li and K. Ren, "Modeling load redistribution attacks in power systems", IEEE Trans. Smart Grid, vol. 2, no. 2, pp. 382-390, Jun. 2011.

[10] A.-H. Mohsenian-Rad and A. Leon-Garcia, "Distributed internet-based load altering attacks against smart power grids", IEEE Trans. Smart Grid, vol. 2, no. 4, pp. 667-674, Dec. 2011.

[11] S. Amini, F. Pasqualetti and H. Mohsenian-Rad, "Dynamic load altering attacks against power system stability: Attack models and protection schemes", IEEE Trans. Smart Grid, vol. 9, no. 4, pp. 2862-2872, Jul. 2018.

[12] Li, Jiangnan, et al. "Towards Adversarial-Resilient Deep Neural Networks for False Data Injection Attack Detection in Power Grids." arXiv preprint arXiv:2102.09057, 2021.

[13] Electric Reliability Council of Texas market (ERCOT). Wind power production: Hourly averaged actual and forecasted values 2012[EB/OL]. 2012. http://www.ercot.com/gridinfo/generation/.

[14] Renewable.ninja 2012. https://www.renewables.ninja